\documentclass[12pt,preprint]{aastex}

\slugcomment{submitted to ApJL}

\shorttitle{Suzaku observation of 1ES1218+304}
\shortauthors{Sato et al.}

\begin{document}


\title{$Suzaku$ observation of TeV blazar the 1ES~1218+304: 
clues on particle acceleration in an extreme TeV blazar}


\author{
   R. Sato\altaffilmark{1},
   J. Kataoka\altaffilmark{2},
   T. Takahashi\altaffilmark{1},
   G. M. Madejski\altaffilmark{3},
   S. R$\ddot{\rm u}$gamer\altaffilmark{4} and
   S. J. Wagner\altaffilmark{5}}
 
 
\altaffiltext{1}{Institute of Space and Astronautical Science/JAXA, Sagamihara, Kanagawa 229-8510, Japan}
 \altaffiltext{2}{Department of Physics, Tokyo Institute of Technology, Meguro-ku, Tokyo 152-8551, Japan}
 \altaffiltext{3}{Stanford Linear Accelerator Center and
Kavli Institute for Particle Astrophysics and Cosmology, Stanford
 University, Stanford, CA 94305, USA}
 \altaffiltext{4}{Universit$\ddot{\rm a}$t W$\ddot{\rm u}$rzburg, Am Hubland, D-97074 W$\ddot{\rm u}$rzburg, Germany}
 \altaffiltext{5}{Landessternwarte, Universit$\ddot{\rm a}$t Heidelberg, K$\ddot{\rm o}$nigstuhl, 69117 Heidelberg, Germany}
 

\begin{abstract}
We observed the TeV blazar 1ES~1218+304 with the X-ray astronomy satellite 
$Suzaku$ in May 2006. 
At the beginning of the two-day continuous observation, 
we detected a large flare in which the 5$-$10 keV flux changed 
by a factor of $\sim$2 on a timescale of 5$\times$10$^4$ s. During the flare,  
the increase in the hard X-ray flux clearly lagged behind that observed 
in the soft X-rays, with the maximum lag of
$2.3\times$10$^4$ s observed between the 0.3$-$1 keV 
and 5$-$10 keV bands. Furthermore we discovered that the temporal profile of 
the flare clearly changes with energy, being more symmetric at higher energies. 
From the spectral fitting of multi-wavelength data assuming a one-zone, 
homogeneous synchrotron self-Compton model, we obtain $B\sim0.047$ G, 
emission region size $R = 3.0\times10^{16}$ cm for an appropriate 
beaming with a Doppler factor of $\delta = 20$. 
This value of $B$ is in good agreement with an independent estimate 
through the model fit to the observed time lag 
ascribing the energy-dependent variability 
to differential acceleration timescale of relativistic electrons 
provided that the gyro-factor $\xi$ is $10^5$.  
%We estimate the magnetic field $B\sim0.049\xi_5$ G where $\xi_5$ is 
%a gyro factor in units of $10^5$. 
\end{abstract}

%% example has been keyed in ApJ style. See the instructions to uthors




\section{Introduction}
 
Blazars are a sub-category of Active Galactic Nuclei where a
relativistic jet pointing close to our line of sight produces Doppler-boosted
emission (e.g., Urry \& Padovani 1995; Ulrich, Maraschi \& Urry 1997).
Generally, their overall spectra have two pronounced continuum components: 
one peaking between IR and X-rays is produced by the synchrotron radiation 
of relativistic electrons, and the other in the $\gamma$-ray regime, 
presumably due to the inverse Compton (IC) emission by the same electrons.
In some cases, $\gamma$-ray emission is seen to 
extend to the TeV range; the X-ray and GeV/TeV $\gamma$-ray bands
correspond to the highest energy ends ($E_{\rm max}$) 
of the synchrotron/IC emission
(e.g., Inoue \& Takahara 1996; Kirk, Rieger \& Mastichiadis 1998).
At these ends, variability is expected to be most pronounced, and
in fact, such large flux variations are observed, on a timescale
of hours to days (e.g., Kataoka et al. 2001; Tanihata et al. 2001) 
or even shorter (minutes scale; Aharonian et al. 2007; Albert et al. 2007).
Using $ASCA$ data, Takahashi et al. (1996) argued the soft X-ray
($<$ 1 keV) variation of Mrk~421, observed to lag behind that of 
the hard X-rays ($\ge$ 2 keV) by $\sim$ 4 ks, may well be ascribed to 
the energy dependence of the synchrotron cooling timescale. More recently,
Kataoka et al. (2000) interpreted an observed soft-lag and spectral
evolution of PKS~2155-304 by a newly developed time-dependent
synchrotron self-Compton (SSC) model.
 
The above $paradigm$ of ``soft-lag'' was questioned, however,
in several aspects. First, intensive X-ray monitoring of
blazars has revealed not only soft lags
but in some cases hard lags (Takahashi et al. 2000) which may be a
manifestation of another process, e.g., energy dependent acceleration.
Second, Edelson et al. (2001) voiced concerns
about reliability of measurement of lags that are smaller
than the orbital periods ($\sim$ 6 ks) of low Earth orbit satellites.
This was refuted by Tanihata et al. (2001) and Zhang et al. (2004) who
showed that, although periodic gaps introduce larger uncertainties than evenly
sampled data, lags on hour-scale cannot be the result of
periodic gaps. A time resolved cross
correlation analysis of uninterrupted Mrk~421 data obtained by XMM-Newton
revealed lags of both signs, changing
on time scales of up to a few 10$^3$ s (Brinkmann et al. 2005).
Hence the situation is very complex and still under debate.
 
In this letter we present new results from the May 2006 $Suzaku$ observation 
of 1ES~1218+304 conducted as part of a multi-wavelength campaign with 
KVA-$Swift$-MAGIC.
1ES~1218+304 is categorized as a high-frequency peaked BL Lac object, 
at a redshift $z = 0.182$ (Veron-Cetty \& Veron 2003). 
It was discovered as a TeV emitter by MAGIC at energies $>100$ GeV
(Albert et al. 2006) and subsequently confirmed by VERITAS (Fortin 2007).
While the detailed multiband analysis is ongoing, 
we focus in this letter on a remarkable X-ray flare observed with $Suzaku$. 
We present temporal and spectral features in $\S$ 3, 
in $\S$ 4 we discuss a physical origin of temporal variability.
% in terms of a toy model incorporating the particle acceleration and cooling. 

\section{Observation and Data Reduction}

1ES~1218+304 was observed with $Suzaku$ (Mitsuda et al. 2007) during
2006 May 20$-$21 UT, yielding a net exposure time of 79.9 ks. 
$Suzaku$ carries four sets of X-ray telescopes (Serlemitsos et al. 2007)
each with a focal-plane X-ray CCD camera (XIS, X-ray Imaging Spectrometer;
Koyama et al. 2007) that is sensitive over the 0.3-12 keV band,
together with a non-imaging Hard X-ray Detector (HXD; Takahashi et al. 2007;
Kokubun et al. 2007), which covers the 10-600 keV energy band with Si PIN
photo-diodes and GSO scintillation detectors.
1ES~1218+304 was focused on the nominal center position of the HXD detector.

For the XIS, we analyzed the screened data, reduced via $Suzaku$ software version 2.0.
The screening was based on the following criteria: 
(1) only ASCA-grade 0,2,3,4,6 events were accumulated, while hot 
and flickering pixels were removed using the CLEANSIS script, 
(2) the time interval after the passage of South Atlantic Anomaly 
is greater than 500 s, (3) the object is at least 5$^\circ$ and 20$^\circ$ 
above the rim of the Earth (ELV) during night and day, respectively. 
In addition, we also select the data with a cutoff rigidity (COR) larger than 6 GV. 
After this screening, the net exposure for good time intervals is 69.4 ks. 
The XIS events were extracted from a circular region with a radius 
of 4.2$^\prime$ centered on the source peak, whereas the background 
was accumulated in an annulus with inner and outer radii of 
5.4$^\prime$ and 7.3$^\prime$, respectively. We checked that the use 
of different source and background regions did not affect the analysis 
results. The response and auxiliary files are produced using 
the analysis tools \textsc{xisrmfgen} and \textsc{xissimarfgen} 
developed by the $Suzaku$ team, 
which are included in the software package HEAsoft version 6.4. 

The HXD/PIN data (version 2.0) were processed with basically the same screening
criteria as those for the XIS, except that ELV\,$\ge$\,5$^\circ$ through
night and day and COR\,$\ge$\,8\,GV.
The HXD/PIN instrumental background spectra were provided by the HXD team 
for each observation (Kokubun et al. 2007; Fukazawa et al. 2006). 
Both the source and background spectra were made with identical good 
time intervals and the exposure was corrected for detector 
deadtime of 6.0\%. We used the response files version 
\textsc{ae\_hxd\_pinhxdnom2\_20070914.rsp}, provided by the HXD team. 

\section{Analysis and Results}

\begin{figure}[t]
\begin{center}
\includegraphics[angle=0,scale=.4]{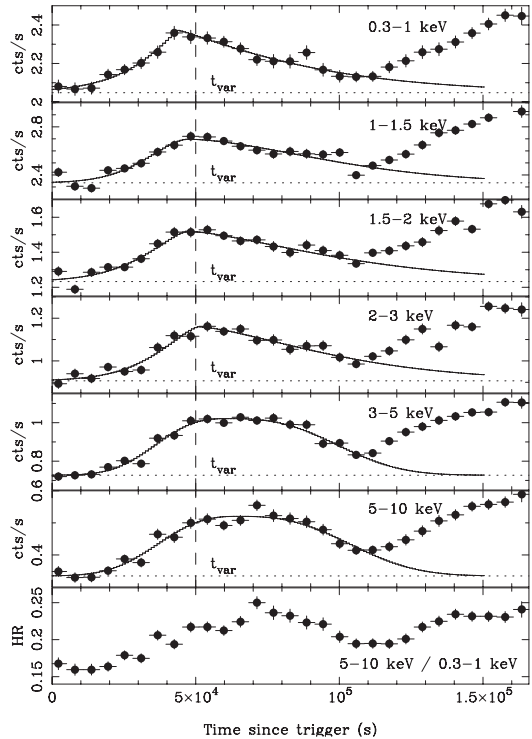}
\caption{Light curves of 1ES~1218+304 observed with $Suzaku$ XISs in 2006 May.
The energy bands are 0.3-1, 1-1.5, 1.5-2, 2-3, 3-5 and 5-10 keV (from the
upper panel), respectively.
The bottom panel shows the HR of count rates, defined as (5-10 keV)/(0.3-1 keV).
The dotted line is the constant offset $C_0$.
The dashed line is the characteristic variability time scale of a flare
$t_{\rm var}$ $\simeq$ 5$\times$10$^4$ s.\label{fig:LC}}
\end{center}
\vspace*{-0.5cm}
\end{figure}

Figure \ref{fig:LC} shows the averaged light curves of the four XISs 
in the six X-ray energy bands. Although we could see variations of 
count rates at some level using HXD/PIN data, 
it was not significant within uncertainties of photon statistics. 
Thus in the following, we concentrate on the temporal variability of 
the XIS data only, below 10 keV. 
The temporal variation  of the hardness ratio (HR) is also shown 
in the bottom panel of Figure \ref{fig:LC}. 
It indicates that the variability in the soft and hard X-ray bands 
are $not$ well synchronized. 

To quantify the different shape of the flare with energy dependent time-lags, 
we fitted the light curves with a function given by Norris (1996) 
after a slight modification of adding a constant offset $C_0$ to mimic the observed light curves:  
\begin{eqnarray*}
I(t) &=& C_0 + C_1 \times \exp [-(|t - t_{\rm peak}/\sigma_{\rm r}|)^{k}] \hspace*{0.2cm} ({\rm for} \hspace*{0.1cm} t \leq t_{\rm peak}),\\
     &=& C_0 + C_1 \times \exp [-(|t - t_{\rm peak}/\sigma_{\rm d}|)^{k}] \hspace*{0.2cm} ({\rm for} \hspace*{0.1cm} t > t_{\rm peak}),\\
\end{eqnarray*}
where $t_{\rm peak}$ is the time of the flare's maximum intensity $C_1$,
$k$ is a measure of pulse sharpness, 
$\sigma_{\rm r}$ and $\sigma_{\rm d}$ are the rise and decay time constants. 
If the light curve is symmetric in time, $\sigma_{\rm r}$ and $\sigma_{\rm d}$
are expected to be equal. 
All the light curves were binned at 2880 s (a half of the orbital period of $Suzaku$) for fitting.
The results of the fittings are given in Table \ref{table:pulsefit}. 
In summary, the observed flare shows the following characteristics:
(1) The flare shape is asymmetric in time ($\sigma_{\rm r}/\sigma_{\rm d}<1$) 
especially in the lower energy band (but note $\sigma_{\rm r}/\sigma_{\rm d}$
$\simeq$ 1 for 5$-$10 keV light curve). 
(2) The flare amplitude defined as $(C_1+C_0)/C_0$ becomes larger as 
the photon energy increases (the 5$-$10 keV flux changed by a factor 
of $\sim$2). 
(3) The rise-time of the flare is almost constant $\sim5\times10^4$ s 
below 2 keV, while it becomes gradually longer at higher energy bands.

Next, we try to evaluate lags of temporal variations in various energy bands. 
Taking into account a wide variety of the flare shape measured at different 
energies, we estimated lags by just comparing the peak-time of the flare 
rather than using other temporal techniques, such as 
the discrete correlation function (DCF; Edelson \& Krolik 1989) or 
the modified mean deviation method (MMD; Hufnagel \& Bregman 1992).
\footnote{Since the DCF quantifies the degree of similarity or correlation 
between two time series as a function of the time-lag, it is not suitable to 
evaluate ``energy-dependent'' profiles, as observed in 1ES~1218+304.}
We compared the peak-time in five lower energy bands to that
determined in the 5$-$10 keV band. 
Apparently, the hard X-ray (5$-$10 keV) peak lagged behind 
that in the soft X-ray (0.3$-$1 keV) by 
(2.3$\pm$0.7)$\times$10$^4$ s. Importantly, this is much larger than 
the orbital period of $Suzaku$ and less affected by artifacts proposed 
in Edelson et al. (2001).

\begin{table}[h]
\small{
\caption{Temporal profiles during the flare of 1ES~1218+304 in various X-ray energy bands.} 
\begin{center}
\vspace*{-0.5cm}
\label{table:pulsefit}
\begin{tabular}{lcccc}
\hline
E (keV) & $t_{\rm peak}$ ($10^4$ s) & $\sigma_{\rm r}/\sigma_{\rm d}$ & $k$ & $(C_1+C_0)/C_0$ \\
\hline
0.3-1 &  4.4$\pm$0.1 & 0.34$\pm$0.14 & 1.2$\pm$0.4 & 1.2$\pm$0.1\\
1-1.5 &  4.7$\pm$0.3 & 0.31$\pm$0.08 & 1.7$\pm$0.5 & 1.2$\pm$0.1\\
1.5-2 &  4.7$\pm$0.3 & 0.31$\pm$0.11 & 1.4$\pm$0.5 & 1.2$\pm$0.1\\
2-3   &  5.1$\pm$0.3 & 0.33$\pm$0.10 & 1.4$\pm$0.4 & 1.3$\pm$0.1\\
3-5   &  6.1$\pm$0.4 & 0.67$\pm$0.12 & 2.7$\pm$0.4 & 1.4$\pm$0.1\\
5-10  &  6.7$\pm$0.7 & 0.84$\pm$0.17 & 2.8$\pm$0.6 & 1.6$\pm$0.1\\
\hline
\end{tabular}
\end{center}
}
\end{table}

The time averaged four XISs and HXD/PIN background subtracted spectra 
were fitted using XSPEC ver.11.3.2, including data within the energy band 0.6$-$50 keV. 
The background of HXD/PIN includes both the instrumental (non X-ray) background 
and the contribution from the cosmic X-ray background (CXB; Gruber et al. 1999). 
Here the form of the CXB was taken as 
$9.0\times10^{-9}(E/3 \hspace{0.1cm} {\rm keV})^{-0.29} \exp(-E/40 \hspace{0.1cm} {\rm keV})$ 
erg cm$^{-2}$ s$^{-1}$ sr$^{-1}$ keV$^{-1}$ and the observed spectrum was simulated assuming 
the PIN detector response to isotropic diffuse emission. 
We first fitted with a single power-law model with Galactic absorption 
$N_{\rm H} = 1.78\times10^{20}$ cm$^{-2}$ (Costamante et al. 2001). 
We obtained the best fit photon index $\Gamma = 2.14\pm0.01$, 
but this model did not represent the spectrum well yielding a reduced $\chi^2$ of 1.23 for 1967 dof. 
We also tried to fit with a broken power-law model with Galactic absorption. 
The photon index below the break energy $E_{\rm brk}$ ($\Gamma_1$) is 
$2.04\pm0.01$ while the index above $E_{\rm brk}$ is $2.17\pm0.01$, 
where $E_{\rm brk}$ ($\Gamma_2$) is $1.42\pm0.05$ keV. 
The flux over 2-10 keV is $\sim 2.0\times10^{-11}$ erg cm$^{-2}$ s$^{-1}$. 
This model gives a better fit with a reduced $\chi^2$ of 1.14 for 1959 dof 
compared to the single power-law model, but $\chi^2$ is still not acceptable. 
Considering the spectral variability, we analyzed the spectrum every 5760 s. 
The power-law indices vary from $2.05\pm0.01$ to $2.22\pm0.01$ during the flare, 
and each segment can be fitted well with a single power-law model or broken power-law 
model with $\chi^2$/dof ranging from 0.94 to 1.09. 

\begin{figure}[t]
\begin{center}
\includegraphics[angle=90,scale=.4]{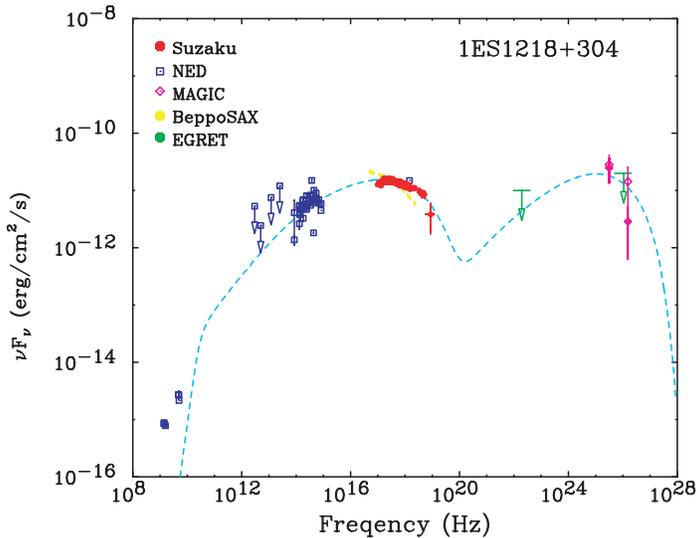}
\caption{Overall SED of 1ES~1218+304.
Filled circles show X-ray data ($Suzaku$; this work). 
For TeV data (Albert et al. 2006; filled diamonds), 
we adopt the correction for the IR EBL by Primack et al. 2001 (open diamonds); 
see also Primack et al. (2005) as well as 
the discussion based on the recent HESS detection of distant TeV blazars (Costamante 2007). 
The other plots are from the NED database. 
The dashed line is a prediction by a one-zone SSC model (Kataoka et al. 2000).
\label{fig:SED}}
\end{center}
\vspace*{-0.5cm}
\end{figure}

Figure \ref{fig:SED} shows the spectral energy distribution (SED) of 
1ES~1218+304 with currently available datasets. 
The TeV data are obtained from Albert et al. (2006), and are corrected for the 
absorption due to the IR Extragalactic 
background light (EBL; see Fig.\ref{fig:SED});  other data are from the 
NED database. Note that the TeV analysis of multi-wavelength campaign data 
is still ongoing and the combined datasets will be investigated in forthcoming paper 
(Stefan et al. in prep). 
As expected from the curved X-ray spectrum with photon index $\Gamma$ around 2 
and $E_{\rm cut} \lesssim 10$ keV, 
the synchrotron emission peaks just around the $Suzaku$ bandpass. 

In order to specify the SED of 1ES~1218+304, 
we applied a one-zone homogeneous SSC model developed in Kataoka et al. (1999). 
Noting that the characteristic variability time scale of the flare is 
$t_{\rm var}$ $\simeq$ 5$\times$10$^4$ s, which is most probably 
determined by the light travel time across the source emitting region 
(see discussion in $\S$ 4), we obtain  $R$ = $c$$t_{\rm var}$$\delta$ 
= 3.0$\times$10$^{16}$ cm for a moderate beaming factor of $\delta$ = 20 
(e.g., Kataoka et al. 1999; 2000 for self-consistent determination of
physical parameters in TeV blazars). With this parameter set, 
the SED of 1ES~1218+304 is fitted with $B=0.047$ G, $s=1.7$, 
$\gamma_{\rm min}$=1, $\gamma_{\rm brk}=8.0\times10^3$ 
and $\gamma_{\rm max}=8.0\times10^5$.
We also note that the energy densities of electrons and fields are 
$u_e=8.3\times10^{-3}$ erg/cm$^3$ and $u_B=8.8\times10^{-5}$ erg/cm$^3$, 
respectively. 
Thus the jet in 1ES~1218+304 is particle dominated, and the 
ratio $u_e$/$u_B$ $\sim$ 100 is well within the range of typical TeV blazars.  

\section{Discussion}

In our observation we detected a large flare during which the hard X-ray 
variation lagged behind that in the soft X-rays, $\sim2.3\times10^4$ s. 
This is completely opposite to a well-known
behavior in which the spectra harden during the phases of rising flux, 
and soften during the phases of falling flux, as has been obtained 
from the past observations. 
In the theoretical context, however, ``hard lag''is actually expected 
especially in the X-ray variability of TeV blazars, 
but has never been observed so clearly before. It has been suggested 
that a hard-lag is observable only at energies closer to the maximum 
electron energy, $\gamma_{\rm max}$ (Kirk, Rieger \&
Mastichiadis 1998), where the acceleration time is  
almost comparable to the cooling time scale of radiating electrons:
$t_{\rm acc}(\gamma_{\rm max})$ $\simeq$ $t_{\rm cool}(\gamma_{\rm max})$.

\begin{figure}[h]
\begin{center}
\includegraphics[angle=0,scale=.35]{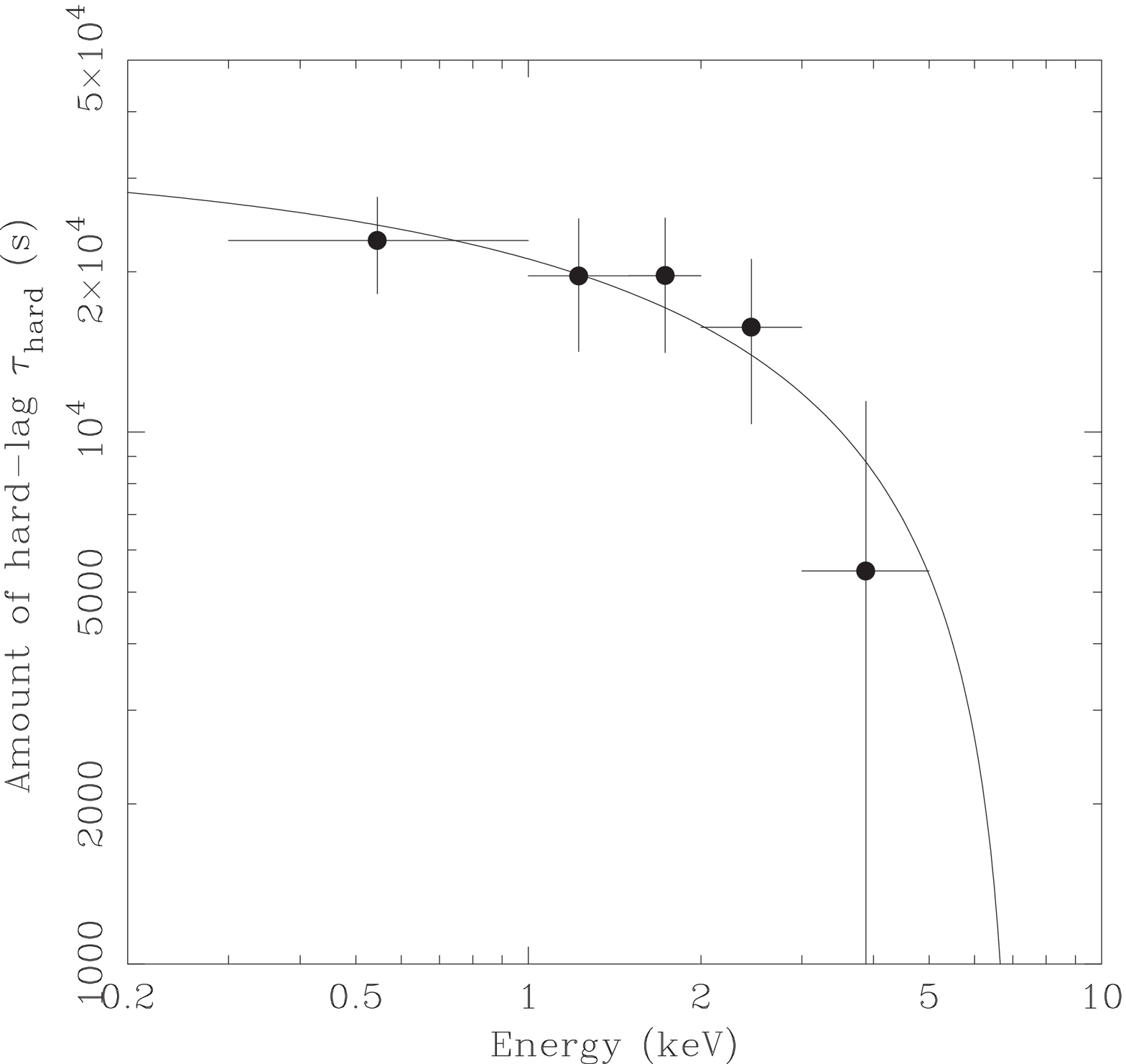}
\includegraphics[angle=0,scale=.35]{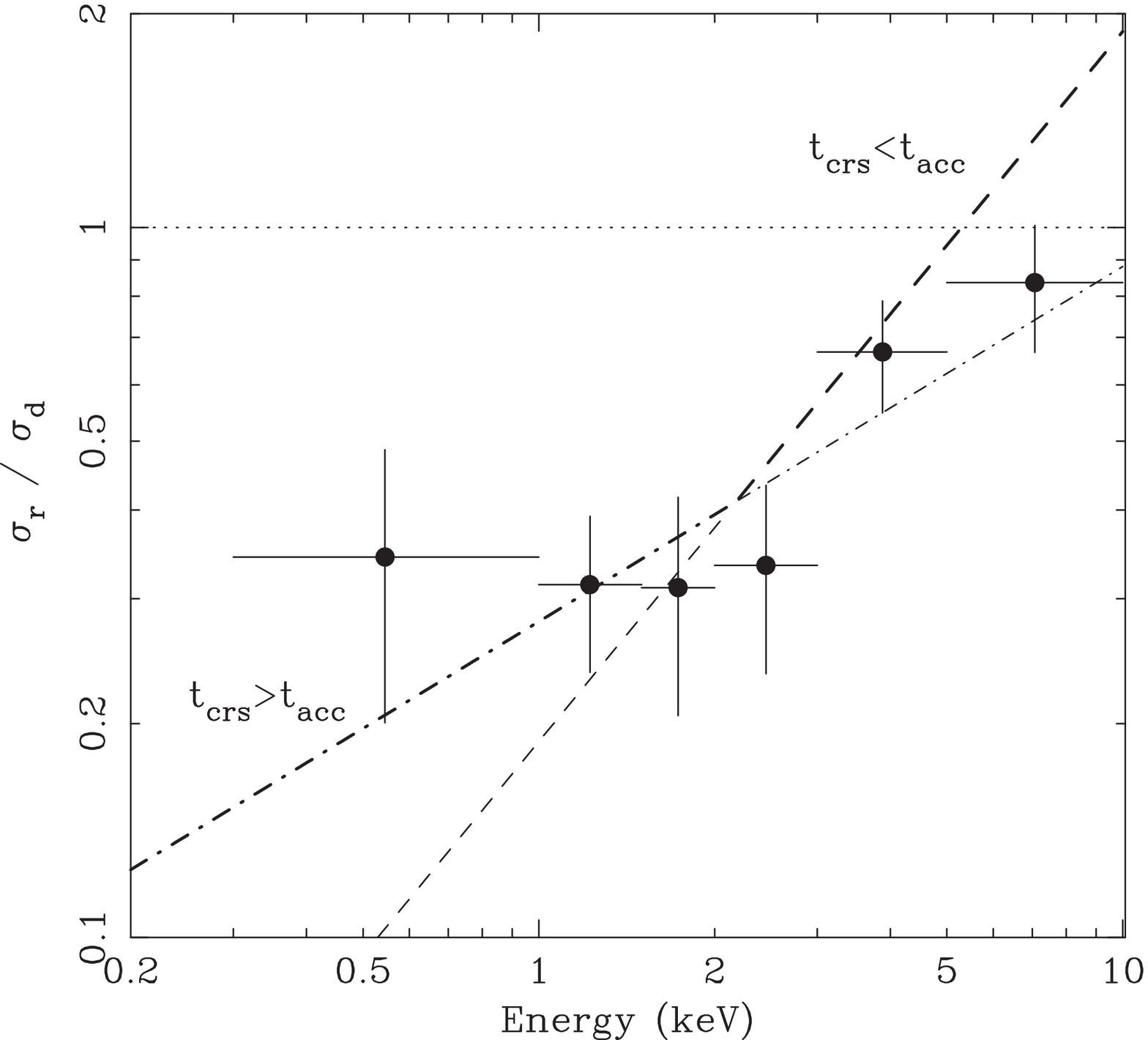}
\caption{$Left$: Time lag of photons of various X-ray energy bands vs
5-10 keV band photons. The solid line corresponds to a fit with
$\tau_{\rm hard} = 9.65\times10^{-2} (1+z)^{3/2} \xi B^{-3/2} \delta^{-3/2} (7.1^{1/2} - E_{\rm low}^{1/2})$ where $\delta$ is fixed to 20.0.
$Right$: Energy dependence of the pulse shape,
defined as the ratio of $\sigma_{\rm r}$ and $\sigma_{\rm d}$.
The dashed line shows the predicted value of $\sigma_{\rm r}/\sigma_{\rm d}$,
calculated from $t_{\rm acc}/t_{\rm cool} = E/E_{\rm max} \sim E/5.3$ keV. 
The dash-dotted line shows the ratio of $t_{\rm crs}/t_{\rm cool}$. \label{fig:lag}}
\end{center}
\end{figure}

It is convenient to express $t_{\rm acc}$ and $t_{\rm cool}$ in terms
of the observed photon energy $E$ (in units of keV).
Noting that the typical synchrotron emission frequency, averaged over pitch angles, 
of an electron with energy $\gamma$$mc^2$ is given by $\nu$ $\sim$
3.7$\times$10$^6$$B$$\gamma^2$ Hz, we obtain; 
\begin{eqnarray*}
t_{\rm acc} (E)  &=& 9.65\times10^{-2} (1+z)^{3/2} \xi B^{-3/2} \delta^{-3/2} E^{1/2} {\rm s},\\
t_{\rm cool} (E) &=& 3.04\times10^{+3} (1+z)^{1/2} B^{-3/2} \delta^{-1/2} E^{-1/2} {\rm s},
\end{eqnarray*}
where $z$ is the redshift, $B$ is the magnetic field strength, $\xi$ is the ``gyro-factor'' 
which can be identified with the ratio of energy in an ordered magnetic 
field to that in a turbulent magnetic field ($\xi$ = 1 for the Bohm limit;  
see, e.g.,  Inoue \& Takahara 1996), and $\delta$ is the beaming factor.  
Note that for lower energy photons, $t_{\rm acc}(E)$ is always shorter 
than $t_{\rm cool}(E)$ because  higher energy electrons need longer time to 
be accelerated ($t_{\rm acc}(\gamma)$ $\propto$ $\gamma$) but cool rapidly 
($t_{\rm cool}(\gamma)$ $\propto$ $\gamma^{-1}$).  This energy
dependence of acceleration/cooling time-scales may qualitatively 
explain the observed characteristics of the X-ray light curves of 1ES~1218+304. 
It is thus interesting to consider a simple toy model in which the
rise time of the flare is primarily controlled by the acceleration time 
of the electrons corresponding to observed photon energies, while the fall time
of the flare is due to the synchrotron cooling time scale. In this
model, the amount of ``hard-lag'', $\tau_{\rm hard}$, is simply due to
the difference of $t_{\rm acc}$, and independent of the energy dependence 
of $t_{\rm cool}$:  
\begin{eqnarray*}
\tau_{\rm hard} &=& t_{\rm acc} (E_{\rm hi}) - t_{\rm acc} (E_{\rm low}) \\
&\sim& 9.65\times10^{-2} (1+z)^{3/2} \xi B^{-3/2} \delta^{-3/2} (E_{\rm hi}^{1/2} 
- E_{\rm low}^{1/2}) \hspace{3mm}{\rm s},\\
\end{eqnarray*}
where $E_{\rm low}$ and $E_{\rm hi}$ are the lower and higher 
X-ray photon energies to which the time-lag is observed.
Here we took $E_{\rm low / hi}$ to be the logarithmic mean energy 
in the observation energy bandpass. 
The result of the model fit to the observed $\tau_{\rm hard}$ is shown 
in Figure \ref{fig:lag} ($left$).  

Assuming a beaming factor $\delta=20$ from multiband spectral fitting 
(see $\S$ 3), the best fit parameter of the magnetic field $B$ can be 
written as $\sim 0.049 \xi_5$ G, where $\xi_5$ is the ``gyro-factor''
in units of $10^5$. 
Thus, in order to have the $B$ field required in the acceleration 
region consistent with that derived from the SED fitting, 
we infer $\xi \sim 10^5$.  Such high value of $\xi$ is in 
fact consistent with that inferred by Inoue \& Takahara (1996) 
for other blazars.
With these parameters, the maximum synchrotron 
radiation energy $E_{\rm max}$, corresponding to $\gamma_{\rm max}$, 
is expected to be $\sim5.3$ keV. 
Hence, the above toy model qualitatively well represents the observed
spectral/temporal features of 1ES~1218+304, in particular:
(1) the synchrotron component peaks around the $Suzaku$ XIS energy band 
in the multiband spectrum (Figure \ref{fig:SED}) and (2) the observed 
light curve is symmetric in shape when measured at the high energy band, 
while being ``asymmetric'' (i.e., fall time longer than the rise time) at the lower energy band.  
Figure \ref{fig:lag} ($right$) compares the energy dependence of observed and
modeled flare shapes, defined as the ratio of rise and decay
time-scales,  $\sigma_{\rm r}$/$\sigma_{\rm d}$. 
The dashed line shows the model prediction from 
$\sigma_{\rm r}$/$\sigma_{\rm d}$ $\simeq$  $t_{\rm acc}/t_{\rm cool}$ =
$(E/E_{\rm max})^{1/2}/(E/E_{\rm max})^{-1/2} \sim E/5.3$ keV.
Although the general trend is well reproduced, Figure \ref{fig:lag} indicates
that the observed rise time may have a bit longer
time scale than expected from the model. The most natural interpretation for
this is the smoothing of rapid variability by the source light
crossing time scale $t_{\rm crs}$ (e.g., Chiaberge \& Ghisellini 1999; 
Kataoka et al. 2000). 
Hence if the acceleration time scale is shorter than the 
source crossing time, we expect $t_{\rm crs}$ to smooth out $t_{\rm acc}$.
The dash-dotted line in Fig. \ref{fig:lag} ($right$) shows the ratio of the time scales 
of $t_{\rm crs}$/$t_{\rm cool}$.
we can see that $t_{\rm crs}$ is longer than $t_{\rm acc}$ below $\sim2$ keV, 
but comparable or shorter above $\sim2$ keV.
As a result, for 1ES~1218+304 it seems reasonable that the rise time of the flare 
is primary determined by the acceleration time of the electrons, while 
the fall time of the flare is due to the synchrotron cooling time scale. 
Only the rise time of the flare in the lower energy bands are dominated 
by the light crossing time $t_{\rm crs}$, 
but further studies using more data are necessary to confirm this model. 

\acknowledgments
We are grateful to Dr. L. Costamante for kindly providing us with the numerical 
values of the EBL effects on TeV gamma-ray spectra, and Dr. M. Sikora for many 
fruitful discussions. This work was supported, in part, by a Department of 
Energy contract to SLAC no. DE-AC3-76SF00515.

%% After the acknowledgments stion, use the following syntax and the
%% in parentheses, after the keyword, but they will not be verified.

%% There should be a \stion command for each appendix. Mark appendix
%% substions with the same markup you use in the main body of the paper.



\begin{thebibliography}{}
\bibitem[Alb06]{Alb06}
Albert, J., et al. 2006, \apjl, 642, L119
\bibitem[Alb07]{Alb07}
Albert, J., et al. 2007, \apj, 669, 862
\bibitem[Aha07]{Aha07}
Aharonian, F. et al., 2007, \apjl, 664, L71
\bibitem[Bri01]{Bri01}
Brinkmann, W., Papadakis, I. E., Raeth, C., Mimica, P., \& Haberl, F.,
	       2005, A\&A, 443, 397
\bibitem[Chi99]{Chi99}
Chiaberge, M., \& Ghisellini, G. 1999, \mnras, 306, 551
\bibitem[Cos01]{Cos01}
Costamante, L., et al. 2001, \aap, 371, 512
\bibitem[Cos07]{Cos07}
Costamante, L., 2007, Ap\&SS, 309, 487 (astro-ph/0612709v1)
\bibitem[Ede89]{Ede89}
Edelson, R.~A., \& Krolik, J.~H. 1989, proceedings of the International Astronomical Union Symposium, 134, 96
\bibitem[Ede01]{Ede01}
Edelson, R., Griffiths, G., Markowitz, A., Sembay, S., Turner, M.~J.~L., 
\& Warwick, R. 2001, \apj, 554, 274
\bibitem[For07]{For07}
Fortin, P. 2007, astro-ph/07093657
\bibitem[Fuk06]{Fuk06}
Fukazawa, Y., et al. 2006, Proc.SPIE, 6266, 75
\bibitem[Gru99]{Gru99}
Gruber, D. E., Matteson, J. L., Peterson, L. E., \& Jung, G. V. 1999, \apj, 520, 124
\bibitem[Huf92]{Huf92}
Hufnagel, B. R., \& Bregman, J. N. 1992, \apj, 386, 473
\bibitem[Ino96]{Ino96}
Inoue, S., \& Takahara, F. 1996, \apj, 463, 555
\bibitem[Kat00]{Kat00}
Kataoka, J., et al. 1999, \apj, 514, 138
\bibitem[Kat00]{Kat00}
Kataoka, J. 2000, Ph.D thesis, Univ. Tokyo (http://www.hp.phys.titech.ac.jp/kataoka/paperJK00-thesis.pdf)
\bibitem[Kat00]{Kat00}
Kataoka, J., et al. 2000, \apj, 528, 243
\bibitem[Kat01]{Kat01}
Kataoka, J., et al. 2001, \apj, 560, 659
\bibitem[Kir98]{Kir98}
Kirk, J.~G., Rieger, F.~M, \& Mastichiadis, A. 1998, \aap, 333, 452
\bibitem[Kok07]{Kok07}
Kokubun, M. et al. 2007, PASJ, 59, S53
\bibitem[Koy07]{Koy07}
Koyama, K. et al. 2007, PASJ, 59, S23
\bibitem[Mit07]{Mit07}
Mitsuda, K. et al. 2007, PASJ, 59, S1
%\bibitem[Muk07]{Muk07}
%Mukherjee, R. et al. 2007, (astro-ph/0710417v1)
\bibitem[Nor96]{Nor96}
Norris, J.~P., et al. 1996, \apj, 459, 393
\bibitem[Pri01]{Pri01}
Primack, J. R., Somerville, E.~S., Bullock, J.~S. \& Devriendt, J.~E.~G. 
2001, AIP Conf. Proc., 558, 463
\bibitem[Pri05]{Pri05}
Primack, J. R., Bullock, J.~S., \& Somerville, E.~S. 2005, AIP Conf. Proc., 745, 23
\bibitem[Ser07]{Ser07}
Serlemitsos, P. J., et al. 2007, PASJ, 59, S9
\bibitem[Ste08]{Ste08}
Stefan, R., et al. 2008, in prep
\bibitem[Tak96]{Tak96}
Takahashi, T. et al. 1996, \apjl, 470, L89
\bibitem[Tak00]{Tak00}
Takahashi, T. et al. 2000, \apjl, 542, L105
\bibitem[Tak07]{Tak07}
Takahashi, T. et al. 2007, PASJ, 59, S35
\bibitem[Tan01]{Tan01}
Tanihata, C. et al. 2001, \apj, 563, 569
\bibitem[Ulr97]{Ulr97}
Ulrich, M-H., Maraschi, L., \& Urry, C. M. 1997, ARA\&A, 35, 445
\bibitem[Urr95]{Urr95}
Urry, C. M., \& Padovani, P. 1995, PASP, 107, 803
\bibitem[Ver03]{Ver03}
Veron-Cetty, M. P. \& Veron, P. 2003, \aap, 412, 339
\bibitem[Zha04]{Zha04}
Zhang, Y. H., Cagnoni, I., Treves, A., Celotti, A., \& Maraschi, L. 2004, \apj, 605, 98
\end{thebibliography}
\end{document}